\documentstyle[12pt,multibox]{article}
\newcommand{\bb}{\begin{equation}}
\newcommand{\ee}{\end{equation}}
\newcommand{\bqn}{\begin{eqnarray}}
\newcommand{\eqn}{\end{eqnarray}}

\renewcommand{\theequation}{\thesection.\arabic{equation}}
\def\d{\delta}
\def\f{\phi}               
\def\g{\gamma}
   
\def\m{\mu}
\def\n{\nu}
  
\def\D{\Delta}

\def\pa{\partial}                              

\def\ck{{\cal K}}
\def\cl{{\cal L}}

\def\co{{\cal O}}

\def\be{\begin{equation}}       \def\eq{\begin{equation}}
\def\ee{\end{equation}}         \def\eqe{\end{equation}}

\def\bea{\begin{eqnarray}}      \def\eqa{\begin{eqnarray}}
\def\ena{\end{eqnarray}}        \def\eea{\end{eqnarray}}
                                \def\eqae{\end{eqnarray}}
\newcommand{\vs}[1]{\vspace{#1 mm}}

\begin{document}
\begin{titlepage}

\begin{flushright}
ESI 780 (1999) \\
ULB-TH-99/24 \\
FTUV/99-66 \\
CERN-TH/99-320\\
MPI/PhT-97-48\\
SPIN-1999/24\\
hep-th/9910201
\end{flushright}
\begin{center}
\vs{10}
{\Large {\bf Cohomological analysis\\ of gauged-fixed gauge theories}}
\end{center}
\vs{3}
\begin{center}
{\large
Glenn Barnich$^{a,b}$, Marc Henneaux$^{b,c}$, Tobias Hurth$^{d,e}$ 
\\
and Kostas Skenderis$^f$}
\end{center}
\vs{5}
\begin{center}{\sl
$^a$ Departament de F\'{\i}sica Te\`orica, Universitat de Val\`encia,\\
E-46100~Burjassot (Val\`encia), Spain.\\[1.5ex]

$^b$ Physique Th\'eorique et Math\'ematique, Universit\'e Libre de
Bruxelles,\\
Campus Plaine C.P. 231, B--1050 Brussels, Belgium\\[1.5ex]

$^c$ Centro de Estudios Cient\'\i ficos de Santiago,\\
Casilla 16443, Santiago 9, Chile\\[1.5ex]

$^d$ CERN, Theory Division,\\
CH-1211 Geneva 23, Switzerland\\[1.5ex] 

$^e$ Max-Planck-Institute for Physics, Werner-Heisenberg-Institute, \\
F\"ohringer Ring 6, D-80805  Munich, Germany \\[1.5ex]

$^f$ Spinoza Insitute, University of Utrecht, Leuvenlaan 4, \\
3584  CE Utrecht, The Netherlands

}\end{center}

\vs{5}

\begin{abstract}
The relation between the gauge-invariant local 
BRST cohomology involving the antifields and the gauge-fixed BRST
cohomology is clarified. It is shown in particular
that the cocycle conditions become equivalent once it is imposed,
on the gauge-fixed side, that the BRST cocycles should yield
deformations that preserve the nilpotency of the (gauge-fixed)
BRST differential. This shows that the restrictions imposed
on local counterterms by the Quantum Noether condition 
in the Epstein--Glaser construction of gauge theories 
are equivalent to the restrictions imposed by BRST invariance
on local counterterms
in the standard Lagrangian approach.
\end{abstract}
\vfill
\vskip 0.2cm

\noindent
\footnotesize{{\tt *
gbarnich@ulb.ac.be, henneaux@ulb.ac.be, tobias.hurth@cern.ch,
K.Skenderis@phys.uu.nl}}
\end{titlepage}

\section{Introduction}
\setcounter{equation}{0}
\setcounter{theorem}{0}

In the BRST approach to perturbative gauge theories \cite{BRS,T},
the possible counterterms are restricted by Ward--Slavnov--Taylor
identities \cite{Ward, Slavnov, Taylor}, which have
a cohomological interpretation.
If one follows the path-integral approach and takes into account
the renormalization of the BRST symmetry \`a la Zinn-Justin \cite{Zinn}
by introducing sources coupled to the BRST variation of the fields
and the ghosts, it may be shown \cite{Zinn,BRS,KG,Dixon,ItzZu}
that the counterterms must fulfil
the BRST invariance condition
\bb
sA = 0,
\label{cocy1}
\ee
where $s$ is the BRST differential acting in the space of fields,
ghosts and associated sources (``antifields" \cite{BV}).  The
counterterms are local, so $A$ in (\ref{cocy1}) is given by
the integral of a local $n$-form $a$, in terms of which the
BRST invariance condition becomes
\bb
sa + db = 0,
\label{cocy2}
\ee
for some $(n-1)$-form $b$.  Following an initial investigation
by Joglekar and Lee \cite{JS}, the general solution of
(\ref{cocy2}) for Yang--Mills gauge models
has been determined in \cite{BBH2}, where it
was shown that up to trivial terms of the
form $sc+de$, the counterterm $a$ in (\ref{cocy2})
is equal to a strictly gauge-invariant operator, plus
Chern--Simons terms in odd space-time dimensions (in
the absence of $U(1)$ factors for which there are
further solutions \cite{Becchi}, also
dealt with in \cite{BBH2}).  This guarantees
renormalizability of the theory in the ``modern sense"
\cite{GW} in any number of spacetime dimensions, and in the standard
power-counting sense in 4 dimensions.

If one follows instead the operator formalism and
the Quantum Noether method based
on the gauge-fixed BRST formulation \cite{HS1,HS2},
one finds that the counterterms are constrained by
the condition
\bb
\gamma^g a + db \approx  0,
\label{cocy3}
\ee
where $\gamma^g$ is the ``gauge-fixed" BRST differential acting
on the fields and where both $a$ and $b$ involve
only the fields (no antifield).  The symbol $\approx$ means ``equal
when the (gauge-fixed) equations of motion hold".

The question then arises as to whether (\ref{cocy2}) and
(\ref{cocy3}) are equivalent. It may be shown that the antifield and
gauge-fixed
local cohomologies are equivalent \cite{Henn1}, so that
any solution $a$ of $sa=0$ defines a solution $a'$ of $\gamma^g
a' \approx 0$ and vice versa. This is {\em not} true, however,
for the cohomologies modulo $d$ \cite{Henn2}.
In particular, there are solutions of (\ref{cocy3}) that have {\em no}
analogue in the antifield cohomology and which, therefore, do
{\em not}  correspond to an integrated, gauge-invariant operator.  An
example is given by the Curci--Ferrari mass term \cite{CurciF}
\bb
-\frac{1}{2} A^a_\mu A^\mu_a + \bar{C}_a C^a,
\label{CFmass}
\ee
which is a solution of (\ref{cocy3}) in the gauge where
the equation of motion for the auxiliary $b$-field is
$b_a + \partial_\mu A_a^\mu - \frac{1}{2} \bar{C}_b
f^b_{ac}  C^c =0$,
but which does not define an integrated gauge-invariant
operator.
The properties of (\ref{CFmass}) have been studied 
in \cite{Ojima,Kostasetal,Hurth1,Dragon,Hurth2,Brandt}.
Thus, (\ref{cocy2}) and (\ref{cocy3}) are in general not
equivalent\footnote{We keep here the auxiliary $b$-field, but 
similar considerations can be made if one eliminates the
auxiliary fields, since the gauge-fixed BRST cohomological
groups are invariant under such an elimination.}.

If, however, the cocycle condition (\ref{cocy3}) is
supplemented by the requirement of nilpotency 
of the deformed BRST differential (which is required
if we want the theory to be unitary)
the Curci--Ferrari mass term is excluded.
It is the purpose of this letter to show that,
quite generally, the gauged-fixed cocycle condition (\ref{cocy3}),
supplemented by the requirement that the deformation generated by
the permissible counterterms should preserve
(on-shell) nilpotency of the BRST symmetry, is equivalent to the
antifield cocycle condition (\ref{cocy2}), which controls the
counterterms in the Zinn-Justin approach. 

This letter is organized as follows. In the next section, we recall some 
salient properties of the gauge-fixed action. The equivalence 
of the two cocycle conditions is shown in section 3, while
a discussion of trivial solutions is presented in section 4. 
In section 5, we review the analysis of counterterms in the Quantum
Noether method and show how nilpotency of the
deformed BRST differential arises in
that context. Finally, in an appendix 
we present an analysis of the relation between the antifield and the weak 
gauged-fixed cohomology using methods of homological algebra.

\section{Gauge-fixed action\label{sec2}}
\setcounter{equation}{0}
\setcounter{theorem}{0}

The starting point is the solution $S[\phi, \phi^*]$ of the master
equation
\bb
(S,S) \equiv 2 \frac{\delta^R S} {\delta \phi^A} \frac{\delta^L S} {\delta
\phi^*_A} \equiv
- 2 \frac{\delta^R S}{\delta \phi^*_A}\frac{\delta^L S}{\delta
\phi^A}
 = 0.
\ee
We use DeWitt's condensed notations.  The solution $S$ is a local
functional, as are all functionals without free indices
occurring below.  The ``fields" $\phi^A$ include the original
fields, the ghosts, as well as the auxiliary fields and the antighosts
of the non-minimal sector.  We assume that
the canonical transformation necessary for gauge-fixing has already
been  performed, so that the gauge-fixed action is simply obtained by
setting the antifields equal to zero, $S^g[\phi] = S[\phi, \phi^*=0]$.
The gauge-fixed BRST differential $\gamma^g$ is defined
by
\bb
\gamma^g \phi^A = -\frac{\delta^R S}{\delta \phi^*_A} \mid_{\phi^*=0},
\label{gammaDef}
\ee
where right and left derivatives are defined by
$\delta F = (\delta^R F/\delta z^\alpha )\delta z^\alpha =
\delta z^\alpha (\delta^L F/\delta z^\alpha )$.  We use the conventions
of \cite{Henn1}, but the derivations are taken to act
from the left (so $sF = (S,F)$ etc.).
The transformation generated by $\gamma^g$ leaves the gauge-fixed
action invariant because of the master equation.  As a result,
the functional derivatives of $S^g$ transform into themselves:
\bb
\gamma^g \frac{\delta^R S^g}{\delta \phi^A} =
- \frac{\delta^R S^g}{\delta \phi^B} \frac{\delta^R}{\delta \phi^A}
\big( \frac{\delta^L S}{\delta \phi^*_B}\big);
\label{gammaEqn}
\ee
here and below,
it is understood that $\phi^*$ is set equal to zero after the
second derivatives have been computed.
The gauge-fixed BRST differential is weakly nilpotent,
\bb
(\gamma^g)^2 \phi^A = \frac{\delta^R S^g}{\delta \phi^B}
\frac{\delta^L}{\delta \phi^*_B} \big( \frac{\delta^R S}{\delta
\phi^*_A}\big).
\label{gamma2}
\ee
Both (\ref{gammaEqn}) and (\ref{gamma2}) are direct consequences of the
definition (\ref{gammaDef}) and the master equation.

The BRST differential in the space of the fields and the
antifields is defined by
\bb
s F = (S, F)
\ee
for any $F(\phi, \phi^*)$.  It is related to $\gamma^g$ as
$s \phi^A = \gamma^g \phi^A + \hbox{antifield-dependent terms}$
or, which is the same,
$\gamma^g \phi^A = s \phi^A \mid_{\phi^*=0}$.
It is strictly nilpotent, $s^2 = 0$.
It will be useful in the sequel to give a special name
to the terms linear in $\phi^*$ in the expansion of $s \phi^A$,
\bb
s \phi^A = \gamma^g \phi^A + \lambda^g \phi^A + O((\phi^*)^2),
\label{sphi}
\ee
with
\begin{eqnarray}
\lambda^g \phi^A &=& (S, \phi^A)\mid_{\hbox{linear in $\phi^*$}}
\nonumber \\
&=& - \phi^*_B \frac{\delta^L}{\delta \phi^*_B} \big( \frac{\delta^R
S}{\delta
\phi^*_A}\big).
\end{eqnarray}

The action of $s$ on the antifields can also be expanded in powers of
the
antifields.  One has
$s \phi^*_A = \delta^g \phi^*_A + \gamma^g \phi^*_A + O((\phi^*)^2)$
where $\delta^g$ is the Koszul differential associated with the
gauge-fixed stationary surface \cite{Henn1},
\bb
\delta^g \phi^*_A = (S, \phi^*_A)\mid_{\phi^*=0} =
\frac{\delta^R S^g}{\delta \phi^A}
\ee
and where
\bb
\gamma^g \phi^*_A = (S, \phi^*_A)\mid_{\hbox{linear in $\phi^*$}}
= \phi^*_B \frac{\delta^R}{\delta \phi^A}
\big( \frac{\delta^L S}{\delta \phi^*_B}\big).
\ee
One easily verifies
the relations $(\delta^g)^2 = 0$,
$\delta^g \gamma^g + \gamma^g \delta^g = 0$,
$(\delta^g\lambda^g + \lambda^g \delta^g) \phi^A
+ (\gamma^g)^2 \phi^A =0$ from the definitions of the
derivations $\delta^g$, $\gamma^g$ and $\lambda^g$.  These
relations are actually the first ones to arise
in the expansion of $s^2= 0$ in powers of the antifields.

The canonical transformation appropriate to gauge-fixing
does not modify the cohomology of $s$ neither in the space of local
functions nor in the space of local functionals, because it is
just a change of variables.  So, in the case
of Yang--Mills theory, the cohomology group $H^0(s, {\cal F})$ of the
BRST differential in the space ${\cal F}$ of local functionals is still
given by the analysis of \cite{BBH2}.
In $H^0(s, {\cal F})$, the superscript $0$ is the total ghost
number.
Note, however, that the expansion
$s = \delta^g + \gamma^g + \lambda^g + \cdots$
is not the standard expansion arising prior to gauge-fixing, 
since the degree involved here
is the total antifield number that gives equal weight to
each antifield, irrespective of its ``antighost" number.
This is why it is the Koszul resolution associated with
the gauge-fixed stationary surface that arises in the present
analysis, and not
the Koszul--Tate resolution associated with the gauge-invariant
equations of motion.

Since the equations of motion following from the
gauge-fixed action have no gauge invariance (by assumption), one
may invoke the general results of \cite{BBH1,Vino,Bryant} to
assert that
\bb
H_k(\delta^g, {\cal F}) = 0 , \quad k \geq 2
\label{acycl}
\ee
where $k$ is the total antifield number used in the above
expansions.  In words:  any local functional $A \in {\cal F}$
($A = \int a$) that  solves $\delta^g A = 0$ ($\delta^g a + db =0$)
and is at least quadratic in the antifields has the form
$A = \delta^g C$ ($a = \delta^g c + dm$).

We have of course $H_k(\delta^g, {\cal L}) = 0$ for $k \geq 1$ in the
space ${\cal L}$ of local functions \cite{Henn1}, but we shall need the
version valid for local functionals below.  This is a direct consequence
of Theorem 8.3 or 10.1 of \cite{BBH1}, which states that there
can be no non-trivial higher-order conservation laws for an action
having
no gauge symmetries.  This theorem is also known as the  ``Vinogradov
two-line theorem".  Because higher-order conservation
laws and elements of $H_k(\delta^g, {\cal F})$ (also
denoted $H_k(\delta^g \vert d)$) are in bijection, the
property (\ref{acycl}) follows.  In general, however, the homological
group $H_1(\delta^g, {\cal F}) $ does not vanish (even though
$H_1(\delta^g, {\cal L}) = 0$ in the space of local functions)
and is related to the
global symmetries of the gauge-fixed action \cite{BBH1}.

\section{Reconstruction Theorem}
\setcounter{equation}{0}
\setcounter{theorem}{0}

We now have all the required tools to show that a local
counterterm of the gauge-fixed formalism that preserves
nilpotency defines a local counterterm of the
antifield  Zinn-Justin approach.  That is, the
condition
\bb
\gamma^g A_0 \approx 0
\label{weak}
\ee
for the local functional $A_0[\phi] = \int a_0$
(which implies (\ref{cocy3}) for the integrand $a_0$),
together with the fact that the associated
deformed BRST symmetry $\gamma^g + e \Delta$
{\em should remain weakly nilpotent} (for the new equations
of motion) to $O(e^2)$ in the deformation parameter
$e$,
\bb
(\gamma^g + e \Delta)^2 \approx' O(e^2),
\label{nilpotence}
\ee
determines a local functional cocycle $A[\phi, \phi^*]$
of the antifield cohomology
\bb
s A = 0.
\label{BRST}
\ee
In (\ref{nilpotence}), the symbol $\approx'$ means ``equal when
the deformed equations of motion $\delta^R (S + e A_0)/\delta \phi^A = 0$
hold".  The relationship between $A$ and $A_0$ is
\bb
A = A_0 + A_1 + A_2 + O((\phi^*)^3),
\label{Expansion}
\ee
where $A_1$ (respectively, $A_2$) is linear (respectively, quadratic)
in the antifields.

The above derivation $\Delta$ is the deformation of the
BRST-symmetry and is related to the deformation $A_0$ of the action as
follows.
When one adds $e A_0$ to the gauge-fixed action,
$S^g[\phi] \rightarrow S^g[\phi] + e A_0[\phi]$,
one modifies the gauge-fixed BRST symmetry as
$\gamma^g \rightarrow \gamma^g + e \Delta$
in such a way that
$(\gamma^g + e \Delta)(S^g + e A_0) = \co(e^2)$.
The existence of $\Delta$ is guaranteed by the cocycle condition
(\ref{weak}), which we  can rewrite as
\bb
\gamma^g A_0 + \delta^g A_1 =0
\label{weak'}
\ee
for some local functional $A_1$ linear in the antifields.
We have $\Delta \phi^A = \Delta^A$, $A_1 = \phi^*_A \Delta^A (-1)^{g_A}$,
where $g_A$ is the Grassman parity of $\phi^A$.

As (\ref{Expansion}) shows, the relationship between $A[\phi, \phi^*]$
and $A_0[\phi]$ is that $A[\phi, \phi^*]$ starts like
$A_0[\phi]$ to zeroth order in the antifields.  Thus, the question is
whether any local functional $A_0$ that fulfils both (\ref{weak})
and (\ref{nilpotence}) can be completed by terms of higher
orders in the antifields to yield a (local functional) solution
of (\ref{BRST}).

The converse statement, namely, that
any local functional $A[\phi, \phi^*]$ solution of
(\ref{BRST}) defines, when setting the antifields equal to zero,
a cocycle of the weak cohomology
(\ref{weak}) fulfilling (\ref{nilpotence}), is rather
obvious.  Indeed, if $sA =0$, then
$\gamma^g A_0 \approx 0$ (term independent of the
antifields in $sA =0$).  Furthermore, at the next order,
\bb
\lambda^g A_0 + \gamma^g A_1 + \delta^g A_2 = 0,
\label{nilpotenceBis}
\ee
a relation that is seen to be equivalent to (\ref{nilpotence})
by rephrasing the condition (\ref{nilpotence}) in terms
of $A_0$ and $A_1$.  On the one hand, 
direct calculations yield 
\begin{eqnarray}
\gamma^g A_1 &=& \phi^*_A \Delta^B \frac{\delta^L}{\delta \phi^B}
\big( \frac{\delta^R S}{\delta \phi^*_A} \big) -
\phi^*_A (\gamma^g \Delta^A) \nonumber \\
\lambda^g A_0 &=& \phi^*_A
\frac{\delta^R A_0}{\delta \phi^B} \frac{\delta^L}{\delta \phi^*_B}
\big( \frac{\delta^R S}{\delta \phi^*_A} \big).
\end{eqnarray}
On the other hand, if one replaces the weak equality by a strong
equality in
$(\gamma^g + e \Delta)^2 \phi^A \approx' O(e^2)$,
one gets,
in view of (\ref{gamma2}),
\bb
(\gamma^g + e \Delta)^2 \phi^A =
\big( \frac{\delta^R S^g}{\delta \phi^B}
+ e \frac{\delta^R A_0}{\delta \phi^B} \big)
\big(\frac{\delta^L}{\delta \phi^*_B}
\big( \frac{\delta^R S}{\delta \phi^*_A} \big)
+ e \mu^{AB} \big) +O(e^2)
\ee
for some $\m^{AB}$.
Thus, (\ref{nilpotence}) becomes to order $e$,
\bb
\gamma^g \Delta^A - \Delta^B \frac{\delta^L}{\delta \phi^B}
\big( \frac{\delta^R S}{\delta \phi^*_A} \big) \approx
\frac{\delta^R A_0}{\delta \phi^B} \frac{\delta^L}{\delta \phi^*_B}
\big( \frac{\delta^R S}{\delta \phi^*_A} \big),
\label{cond1}
\ee
which shows that (\ref{nilpotence}) is indeed equivalent to the
statement
that $\gamma^g A_1 + \lambda^g A_0$ vanishes weakly, or which is the
same,
(\ref{nilpotenceBis}).

Accordingly, to each counterterm
of the antifield Zinn-Justin approach corresponds a counterterm of the
BRST-Noether method.

Conversely, given a solution of (\ref{weak}) -- or
(\ref{weak'}) --
which also fulfils (\ref{nilpotence}), the question is whether
one can construct a local functional $A$ that starts like
$A_0 + A_1$ and  is BRST-invariant.
That (\ref{weak}) (or (\ref{weak'})) by itself does not guarantee
the existence of $A$ is illustrated by the Curci-Ferrari mass term
and has been explained in \cite{Henn2}.

The problem arises because the perturbative construction,
yielding successively $A_2$, $A_3$, etc., given the ``initial
data" $A_0$ and $A_1$ along the lines of homological perturbation
theory applied to the antifield formalism
\cite{FischH,Henn0,Henn1}  can be obstructed in the space of local
functionals.
The obstructions are in the homological groups
$H_k( \delta^g, {\cal F})$ (also denoted by $H_k( \delta^g \vert d)$)
The point is that the equations defining the higher-order
terms $A_2$, $A_3$ etc. take the form
\bb
\delta^g A_k = B_{k-1},
\label{construction}
\ee
where the local functional $B_{k-1}$ involves only the lower-order terms
$A_i$
($i<k$) and can be shown to be $\delta^g$-closed.  To infer that
$B_{k-1}$ is exact, one needs either $H_{k-1}( \delta^g, {\cal F}) =0$
or,
if $H_{k-1}( \delta^g, {\cal F})$ does not vanish, additional
information
guaranteeing that $B_{k-1}$ is in the zero class.

As we recalled above, $H_j( \delta^g, {\cal F}) =0$ for $j>1$.
Thus the only obstructions may arise for $k-1 =1$, i.e. for $A_2$.  If
it can be proven that $A_2$ exists, there cannot be any further obstruction
at the next orders, and $A$ also exists.  The strategy of the
construction
of $A$ from $A_0$ and $A_1$ consists, then, in showing that one
avoids the obstruction for $A_2$.  It is here that the condition
(\ref{nilpotence}) is necessary.

The equation (\ref{construction}) for $A_2$ is actually
(\ref{nilpotenceBis}) with
\bb
B_1 = - \lambda^g A_0 - \gamma^g A_1.
\label{B1A2}
\ee
We must show that $B_1$ is $\delta^g$-exact, i.e. that it vanishes
weakly.  But this is guaranteed because
(\ref{nilpotenceBis})
and (\ref{nilpotence}) have been shown to be equivalent, so that
(\ref{nilpotence})
implies (\ref{nilpotenceBis}) or (\ref{B1A2}).
Therefore,
the obstruction for $A_2$ is avoided, as announced.

One may understand the equivalence between (\ref{nilpotence}) and
(\ref{nilpotenceBis}) more directly, in terms of the master equation
itself.  As is known \cite{BaHe}, the elements of $H^0(s, {\cal F})$
can be viewed as consistent, first-order deformations of the master
equation, $S \rightarrow S'=S + e A$, $(S,S)=0 \rightarrow
(S',S') = O(e^2)$.  As we have indicated, given $A_0$, the
obstruction to the construction of $A$ can only occur for $A_2$,
i.e. we  must verify that the term (\ref{nilpotenceBis})
is zero.  But this term is the term linear in the antifields in
the master equation.  So, the absence of obstruction is
equivalent to the statement that $(S',S') 
\mid_{\hbox{linear in $\phi^*$}}$
vanishes, or $((S',S'), \phi^A) \mid_{\phi^* =0} = 0$.
This is precisely the statement that the deformed BRST symmetry
remains nilpotent,
as the Jacobi identity for the antibrackets easily shows.

\section{Trivial Solutions}

The map between the antifield cohomology and the gauged-fixed
cohomology fails to be surjective, since only classes
with representatives fulfilling the extra condition 
(\ref{nilpotenceBis}) are in the image of the map.  The
map fails also to be injective, because there are non-trivial
cocycles of the antifield cohomology that are mapped on
trivial cocycles of the gauged-fixed
cohomology.
This is best seen on a simple
example. Consider electromagnetism with a neutral scalar field $\phi$
and impose the gauge condition
$\partial_\mu A^\mu= \mu \phi$ through the equation of
motion for the auxiliary $b$-field, where $\mu$ is a constant with
dimension $L^{-1}$. With that
gauge choice, the nontrivial
cocycle $\int d^4x\ \phi$ of the gauge-invariant
cohomology becomes trivial in the weak gauge-fixed cohomology
since one has $\phi \approx s \bar{C} + \partial_\mu A^\mu$.
Similar considerations would apply to any function $f(\phi)$
in the gauge $\partial_\mu A^\mu=f(\phi)$.  
Although we will not provide a precise argument,
we note that these mod-$d$ coboundaries of the gauge-fixed cohomology,
which are present in peculiar gauges, are not expected
to be physically trivial. The reason is that
correlation functions of gauge-invariant operators do not change 
in different gauges (for a proof within the EG framework, see 
\cite{HS3}).   

Note that the gauge-fixed action
has a nontrivial
global symmetry acting on the unphysical variables, namely
the shift
$\bar C\rightarrow \bar C + \theta $, where $\theta$ is a constant
Grassmann odd parameter, corresponding to the cohomology class $\int
d^4x\ \bar C^*$.  This phenomenon is precisely related in the appendix to
the  non-injectivity of the above map.

\section{Counterterms in the Quantum Noether method}
\setcounter{equation}{0}

We show in this section how the nilpotency condition arises
in the Quantum Noether method. This method \cite{HS1,HS2}
is a general method for constructing theories with global symmetries
using the Epstein-Glaser (EG) approach to quantum 
field theory. In this approach, which was introduced by Bogoliubov 
and Shirkov \cite{BS} and developed by Epstein 
and Glaser \cite{EG0,EG2}, the 
(perturbative) S-matrix is directly constructed in the 
Fock space of asymptotic fields 
by imposing causality and Poincar\'{e} invariance.
The method can be regarded as an ``inverse'' of the cutting rules:
one builds $n$-point functions by appropriately ``gluing'' together
$m$-point functions ($m<n$). Moreover, this method directly yields
a finite perturbation theory; one avoids UV infinities altogether by 
proper treatment of $n$-point functions as operator-valued
distributions. The coupling constants of the theory, $e$, are 
replaced by tempered test functions $g(x)$ (i.e. smooth functions rapidly 
decreasing at infinity), which switch on the interactions.
The iterative construction of the S-matrix starts by 
giving a number of free fields satisfying (gauged-fixed) 
fields equation (so that there are propagators) and  
the first term, $T_1$, in the perturbative expansion 
of the S-matrix. Ultimately, one is interested in the
theory in which $g(x)$ becomes again constant,
$g(x) \to e$. This is the so-called adiabatic limit. 
We use the convention to still keep $e$ explicit, in 
which case the adiabatic limit is $g(x) \to 1$. 
We work before the adiabatic limit is taken, as the 
latter does not always exist because of physical infrared
singularities. 

Causality and Poincar\'{e} invariance completely fix the S-matrix up to
local terms. The remaining local ambiguity is
further constrained by symmetries.
It is the purpose of our analysis to determine the precise
restrictions imposed on these local terms by Ward identities.
At tree level the local terms are equal to the Lagrangian of the conventional 
approach \cite{HS1}, but new local terms may be introduced at each 
order in perturbation theory. The local terms at loop level correspond to 
the counterterms in the Lagrangian approach,
although their role is not to subtract infinities, as the
perturbative expansion is already finite.
If the form of these local terms remains the same to all orders in 
perturbation theory then the theory is renormalizable.

The Quantum Noether method consists of adding a coupling to the 
Noether current $j_0^\m$ that generates the asymptotic (and hence 
linear) symmetry in the theory and then requiring that this current be 
conserved inside correlation functions. There are a number of 
equivalent ways to present this condition \cite{HS1,HS2}.
Here we follow \cite{HS2}, where the condition was 
formulated in terms of the interacting Noether current. The Ward 
identity, formula (3.1) in \cite{HS2}, contains terms
that vanish in the (naive) adiabatic limit, $g(x) \to 1$.
Their explicit form, which can be found in \cite{HS2},
is not important for the present analysis.
Here we will schematically denote them by $\pa_\m g \tilde{j}^\m$.
Due to these terms the interacting BRST charge
is not conserved before the adiabatic limit is 
taken. For a discussion of the implications of this fact
(and also of other difficulties encountered when attempting to 
construct the interacting BRST charge) we refer to 
\cite{fred}. We note, however, that considerations
involving only currents are sufficient in order  
to derive all consequences of nonlinear symmetries
for time-ordered products. 
The Quantum Noether condition reads
\be \label{QNC1}
\pa_\m^x T[j_0^\m(x) T_1 (x_1) \cdots \cdots T_1(x_n) ] = 
\pa_\m g \tilde{j}^\m.
\ee
Working out the consequences of this condition to all orders, one recovers 
the non-linear structure in a manner similar to the way the
Noether method works in classical field theory \cite{HS1,HS2}.
Further consistency requirements on the theory follow by considering 
multi-current correlation functions. In particular, the 
two-current equation is
\be \label{QNC2}
\pa_\m^x 
T[j_0^\m(x) j_0^\n(y) T_1 (x_1) \cdots T_1(x_n)] = \pa_\m g \tilde{j}^{\m \n},
\ee
where again we have only schematically included terms that vanish in the 
naive adiabatic limit. The explicit form of these terms, as well
as an all-order analysis of (\ref{QNC2}), will be 
presented in \cite{HS3}.

We are interested in gauge theories. In this case the 
relevant symmetry is BRST symmetry. We now present the analysis
of (\ref{QNC1}), (\ref{QNC2}) for this case to first 
non-trivial order. This is sufficient in order to connect with 
the analysis of the preceding sections.
Equation (\ref{QNC1}) at first order yields the following 
condition on $\cl_1=(\hbar/i) T_1$ :
\be \label{1st}
\g^g \cl_1 = \pa_\m \cl^\m_1 + e \D \f^A \ck^{(0)}_{AB} \f^B,
\ee
where $\f^A$ denotes collectively all the fields;
$\g^g= [\int j^0,\cdot]$ generates the asymptotic transformation rules;
$\D \f^A$ is defined by eq. (\ref{1st}).
It was shown in \cite{HS1,HS2} that 
$\D \f^A$  is the next-order symmetry transformation
rule; $\cl^\m_1$ is some local function of $\f^A$ and its first 
derivative $\pa_\m \f^A$, and $\ck^{(0)}_{AB} \f^B$
are the free-field equations.

To work out the consequences of condition (\ref{QNC2}), we first 
note that since $j_0^\m$ is the gauged-fixed BRST current it 
satisfies
\be \label{clas}
\g^g j_0^\m = \pa_\n T_0^{\m \n} + 
J_0^{\m A} \ck^{(0)}_{AB} \f^A,
\ee
where $T_0^{\m \n}$ is antisymmetric in $\m, \n$, 
and $J_0^{\m A}$ may contain derivatives acting on the free-field
equations. 
Equation (\ref{clas}) guarantees that (\ref{QNC2}) is satisfied
at $n=0$ (i.e. no $T_1$ involved). At $n=1$ one finds the following 
condition:
\be
J_0^{\m A} {\d \cl_1 \over \d \f^A} + \g^g j_1^\m + 
\D j_0^\m = \pa_\m T_1^{\m \n} + J_1^{\m A} \ck^{(0)}_{AB} \f^B,
\ee
for some $T_1^{\m \n}$ and $J_1^{\m A}$ (also possibly containing 
derivatives
acting on $\ck^{(0)}_{AB} \f^B$);  
$\d \cl_1/\d \f^A$ is the Euler derivative of $\cl_1$, 
and if $J_0^{\m A}$ contained derivatives in (\ref{clas}) 
they now act on $\d \cl_1/\d \f^A$; $j_1^\m$ arises 
as a local normalization term of the correlation function 
$T[j_0^\m(x_1) T_1(x_2)]$. It was shown in \cite{HS2} that it
is the Noether current that generates the symmetry transformation
rules $\D \f^A$. Combining with (\ref{clas}) we obtain
\be \label{2nd}
(\g^g + e \D) (j_0^\m + e j_1^\m) = \pa_\m (T_0^{\m \n} + e T_1^{\m \n}) 
+ (J_0^{\m A} + e J_1^{\m A}) \ck_{AB}^{(1)} \f^A + O(e^2),
\ee
where $\ck_{AB}^{(1)} \f^A$ are the field equations that follow
from the Lagrangian $\cl_0 + e \cl_1$, where $\cl_0$ generates
the free field equations. 

Conditions (\ref{1st}) and (\ref{2nd}) are equivalent to conditions
(\ref{weak}) and (\ref{nilpotence}) we analysed in section 3.

\section{Conclusions}
In this letter, we have shown that the restrictions imposed 
on counterterms by the Quantum Noether condition 
in the Epstein--Glaser construction of gauge theories
are equivalent to those imposed in the 
Zinn-Justin (``antifield") approach to the renormalization of gauge
theories. The crucial requirement that guarantees the equivalence
of the restrictions on the counterterms (``cocycle conditions")
is the nilpotency of the deformed BRST generator.
We have also analysed how this requirement arises in
the EG approach.  Similar considerations apply
to anomalies. This will be discussed elsewhere \cite{HS3}.

\section*{Acknowledgements}

The authors acknowledge the hospitality of the Erwin Schr\"odinger
International
Institute for Mathematical Physics in Vienna, where this collaboration has
been started.
This work has been partly supported by the ``Actions de
Recherche Concert{\'e}es" of the ``Direction de la Recherche
Scientifique - Communaut{\'e} Fran{\c c}aise de Belgique", by
IISN - Belgium (convention 4.4505.86), by
Proyectos FONDECYT 1970151 and 7960001 (Chile).
TH was supported by the DOE under grant No. 
DE-FG03-92-ER40701 during a visit of the theory group 
at CALTECH where part of this work was done.
KS is supported by the Netherlands Organization for 
Scientific Research (NWO).

\section*{Appendix: Antifield (``canonical")
versus weak gauge-fixed BRST cohomology}
\renewcommand{\theequation}{A.\arabic{equation}}
\setcounter{equation}{0}
\setcounter{theorem}{0}

In this appendix, the general relation in the space of local
functionals ${\cal F}$ between the antifield BRST cohomology computed
before
gauge-fixing and the weak gauge-fixed version is analysed by
using standard tools from homological algebra.

As mentioned in section
\ref{sec2}, the canonical transformation used for gauge-fixing does
not modify the antifield BRST cohomology and we assume that this
transformation has been done. The complete BRST differential $s$ in
canonical form then differs from that in gauge-fixed form only in the 
grading
used for the expansion, called generically ``resolution degree" below.
The grading associated to
the canonical form consists in assigning antighost number $1$ to the
antifelds of the original fields, $2$ to the antifields of the ghosts,
$3$ to the antifields of the ghosts for ghosts, etc.,
while in the gauge-fixed case
the grading consists in assigning antifield number $1$ to
all the antifields.

In both cases, we have an expansion of the form
$s=\delta^\prime+\gamma^\prime+\sum_{k\geq 1}s^\prime_k$,
in the bigraded space $V=\oplus_{k,g}V^g_k$, with
$g\in {\bf Z}$ the ghost number and $k\in {\bf N}$ the resolution
degree. The ghost number of $s$ is $1$, the resolution degree of
$\delta^\prime$, $\gamma^\prime$, $s^\prime_k$ are respectively $-1$,
$0$, $k$.

Let $V_{k\geq n}$ be the space containing only terms
of resolution degree
larger  than $n$: $A\in V_{k\geq n}$ if the expansion of $A$ according
to the resolution degree is $A=A_n+A_{n+1}+\dots$. In particular
$V=V_{k\geq 0}$.

For $n\geq 0$, consider the spaces $H^g(s,V_{k\geq n})$
defined by the cocycle
condition $s(A_n+A_{n+1}+\dots)=0$ and the coboundary condition
$A_n+A_{n+1}+\dots =s(B_n+B_{n+1}+\dots)$.
In particular, $\delta^\prime B_n=0$.
Consider the maps $i_{n}: H^g(s,V_{k\geq
n+1})\longrightarrow H^g(s,V_{k\geq n})$ defined by
$i_{n}[A_{n+1}+A_{n+2}
+\dots ]=[A_{n+1}+A_{n+2}+\dots ]$.
They are well defined because they map cocycles to cocycles and
coboundaries
to coboundaries.
Note that the difference between $H^g(s,V_{k\geq
n+1})$ and ${\rm im}\ i_{n}$ is the coboundary condition: an element
$A=A_{n+1}+A_{n+2}+\dots$, with $s A=0$, is trivial in
${\rm im}\ i_n\subset
H^g(s,V_{k\geq
n})$, if $A=s B$ with $B=B_n+B_{n+1}+\dots$.

For $n\geq 0$, consider the spaces
$H^g_n(\gamma^\prime,H(\delta^\prime,V))$.
The
cocycle condition for an element $[A_n]\in  H^g_n(\gamma^\prime,
H(\delta^\prime,V))$
is $\delta^\prime A_n=0$, $\gamma^\prime A_n+\delta^\prime
A_{n+1}=0$
for some $A_{n+1}$, and the coboundary condition is
$A_n=\gamma^\prime
B_n+\delta^\prime B_{n+1}$, with $\delta^\prime B_n=0$.
Consider the maps $\pi_n: H^g(s,V_{k\geq n})\longrightarrow
H^g_n(\gamma^\prime,H(\delta^\prime,V))$ defined by
$\pi_n[A_n+A_{n+1}+\dots]=[A_n]$. 

Consider finally the maps
$m_n: H_n^g(\gamma^\prime,H(\delta^\prime,V))\longrightarrow
H^{g+1}(s,V_{k\geq n+1})$
defined by $m_n[A_n]=[s (A_n+A_{n+1})]$.
It is straightforward to check that the maps $m_n$ 
are well defined. 

We are now in a position to prove the decomposition:
\begin{equation}
H^g(s,V_{k\geq n})\simeq
{\rm ker}\ m_n\oplus {\rm im}\ i_n.\label{A1dec}
\end{equation}
The proof follows from the isomorphism (as real vector
spaces) $H^g(s,V_{k\geq n})\simeq {\rm
im }\ \pi_n \oplus{\rm ker}\ \pi_n$ and by showing that 
${\rm ker}\ \pi_n={\rm im}\ i_n$ and 
${\rm im}\ \pi_n={\rm ker}\ m_n$. 
{}From (\ref{A1dec}), it then follows that
\begin{eqnarray}
H^{g}(s,V)\simeq {\rm ker}\ m_0
\oplus i_0[H^g(s,V_{k\geq 1})]\nonumber\\
\simeq {\rm ker}\ m_0\oplus i_0[{\rm ker}\ m_1\oplus i_1
[H^g(s,V_{k\geq
2})]]\simeq \dots\nonumber\\
\simeq {\rm ker}\ m_0\bigoplus_{n\geq 1}i_0
\circ\dots\circ i_{n-1}[{\rm ker}\ m_n].
\end{eqnarray}
Note that the isomorphism 
$H^g(s,V_{k\geq n}) \simeq  {\rm im}\ \pi_n \oplus {\rm ker}\ \pi_n$ used
in the proof is non-canonical in the sense that it involves
a choice of supplementary subspace to ${\rm ker}\ \pi_n$.

{\bf Discussion:}
If $V$ is the space of local functions or of horizontal forms, we have
$H_n(\delta^\prime,V)=0$ for $n\geq 1$, and this both in
the canonical and the gauge-fixed form.
It follows that $H^g_n(\gamma^\prime,H(\delta^\prime,V))=0$ and thus
${\rm ker}\
m_n=0$ for $n\geq 1$. Since $H^{g+1}(s,V_{k\geq
1})=0$ it also follows that $m_0=0$ and ${\rm
ker}\ m_0=H^{g}_{0}(\gamma^\prime,H(\delta^\prime,V))$, so that
$H^g(s,V)\simeq H^{g}_{0}(\gamma^\prime,H(\delta^\prime,V))$. This
result has been deduced in \cite{FischH,Henn1}.

If $V$ is the space of local functionals ${\cal F}$, for the canonical
form of the BRST differential (with the cohomologically trivial pairs
of the non-minimal sector eliminated),
there are no fields with negative pure ghost
numbers. This implies that the
antifield number must be larger than  or equal to  ${\rm max}(0,-g)=K$.
Furthermore, if $k>K$, the presence of the ghosts implies
\cite{Henneaux:1991rx} that
$H^{g}_k(\delta,{\cal F})=0$.
This implies, for $g\geq 0$, that $H^{g}_k(\delta,{\cal F})=0$ for $k\geq
1$, hence ${\rm ker}\ m_n=0$, for $n\geq 1$ and $m_0=0$. Again we get
$H^g(s,{\cal F})\simeq H^{g}_{0}(\gamma,H(\delta,{\cal F}))$.

For $g<0$, the only non-vanishing
cohomology group is $H^g_{-g}(\delta,{\cal F})$. This implies
$H^g_n(\gamma,H(\delta,{\cal F}))=0$, for $n\neq -g$,
so that ${\rm ker}\ m_n=0$ for
$n\neq -g$. Furthermore, $H^{g+1}(s,{\cal F}_{k\geq -g+1})=0$, so that
$m_{-g}=0$. Hence $H^g(s,{\cal F})\simeq i_0\circ\dots\circ
i_{-g-1}[H^g_{-g}(\gamma,H(\delta,{\cal F}))]$.
Finally, $H^g_{-g}(\gamma,H(\delta|d))\simeq H^g_{-g}(\delta|d)$, which
follows from $H^{g+1}_{-g}(\delta,{\cal F})=0$, and
$i_{-g-1}=\dots =i_0=1$ at ghost number $-g$, since there are no terms
with antifield number less than $-g$, so that
$H^g(s,{\cal F})\simeq H^g_{-g}(\delta,{\cal F})$, which is the result
obtained in \cite{BBH2}.

As already stated in section \ref{sec2}, in the space of local
functionals for the gauge-fixed form,
$H^g_k(\delta^g,{\cal F})=0$ for $k\geq
2$, with
$H^{g}_1(\delta^g,{\cal F})$
characterizing the non-trivial global symmetries of the gauge-fixed
action (and their associated
Noether currents) for the classical fields, the
ghost fields and the fields of the gauge-fixing sector.
We thus have ${\rm ker}\ m_2={\rm ker}\ m_3=\dots =0$ and $m_1=0$,
implying that
\begin{eqnarray}
H^g(s,{\cal F})\simeq [{\rm ker}\ m_0 \subset
H^g_0(\gamma^g,H(\delta^g,{\cal F}))]\oplus
i_0[H^g_1(\gamma^g,H(\delta^g,{\cal F}))].
\end{eqnarray}
It follows that the canonical antifield BRST cohomology 
$H^g(s,{\cal F})$ is isomorphic
to the direct sum of a subset of the weak gauge-fixed BRST cohomology
$H^g_0(\gamma^g,H(\delta^g,{\cal F}))$
and of a subset of the nontrivial global symmetries of the gauge-fixed 
action.
Since $H^{g+1}(s,{\cal F}_{k\geq 1})\simeq
H^{g+1}_1(\gamma,H(\delta,{\cal F}))$, the condition that
$[a_0]\in{\rm ker}\
m_0$ becomes $s_1 A_0+\gamma A_1=\gamma B_1+\delta B_2$, with $\delta
B_1=0$. This is precisely condition (\ref{nilpotenceBis}) and thus
equivalent to (\ref{nilpotence}).


\begin{thebibliography}{99}

\bibitem{BRS} C. Becchi, A. Rouet and R. Stora, 
``Renormalization of the abelian Higgs-Kibble model'',
{\em Commun. Math. Phys.} {\bf 42} (1975) 127~; 
``Renormalization of gauge theories'',
{\em Ann. Phys.} (NY) {\bf 98} (1976) 287.

\bibitem{T} I.V. Tyutin, ``Gauge invariance in field theory
and statistical mechanics'',
Lebedev preprint FIAN,  n$^0$39 (1975).

\bibitem{Ward} J.C. Ward, ``An identity in Quantum Electrodynamics'',
{\em Phys. Rev.} {\bf 78} (1950) 182.

\bibitem{Slavnov} A.A. Slavnov, 
``Ward identities in gauge theories'',
{\em Theor. Math. Phys.} {\bf 10} (1972) 99. 

\bibitem{Taylor} J.C. Taylor, 
``Ward identities and charge renormalization of the Yang-Mills field'',
{\em Nucl. Phys.} {\bf B33} (1971) 436. 

\bibitem{Zinn} J. Zinn-Justin, ``Renormalisation of gauge
theories'' in
{\em Trends in elementary particle theory}, Lecture notes in Physics
n$^0$37 (Springer, Berlin, 1975)~;  {\em Quantum Field Theory and
Critical Phenomena}, $3^{\rm rd}$ (Edition Clarendon Press,
Oxford, 1996).

\bibitem{KG} H. Kluberg-Stern and J.B. Zuber,
``Ward identities and some clues to the renormalization of gauge
invariant operators'',
{\em Phys. Rev.} {\bf D12} (1975) 467~;
``Renormalization of nonabelian gauge theories in a 
background field gauge: 1. Green functions'',
{\em Phys. Rev.} {\bf D12} (1975) 482~;
``Renormalization of nonabelian gauge theories in a background field 
gauge. 2. Gauge invariant operators'',
{\em Phys. Rev.} {\bf D12} (1975) 3159.

\bibitem{Dixon} J.A. Dixon, 
``Calculation of BRS cohomology with spectral sequences'',
{\em Commun. Math. Phys.} {\bf 139} (1991)
495.

\bibitem{ItzZu} C. Itzykson and J.B. Zuber, {\it Quantum Field Theory}
 (Mc Graw-Hill, New York, 1980).

\bibitem{BV}  I.A. Batalin and G.A. Vilkovisky, ``Gauge algebra and
quantization'', {\em Phys. Lett.}
{\bf B102} (1981) 27~; ``Quantization of gauge theories with linearly
dependent generators'', 
{\em Phys. Rev.} {\bf D28} (1983) 2567~; 
[erratum: 
{\em Phys. Rev.} {\bf D30}
(1984) 508].

\bibitem{JS} S.D. Joglekar and B.W. Lee, ``General theory of
renormalization of gauge invariant operators'', {\em Ann. Phys.} (NY)
{\bf 97} (1976) 160.

\bibitem{BBH2} G. Barnich and M. Henneaux, ``Renormalization of gauge
invariant operators and anomalies in Yang-Mills theory'', 
{\em Phys. Rev. Lett.}
{\bf 72} (1994) 1588~, hep-th/9312206
; G. Barnich, F. Brandt and M. Henneaux, ``Local
BRST cohomology in the antifield formalism. II. Application to
Yang-Mills theory'', {\em
Commun. Math. Phys.} {\bf 174} (1995) 93, hep-th/9405194.

\bibitem{Becchi} G. Bandelloni, A. Blasi, C. Becchi and
R. Collina, ``Nonsemisimple gauge models: 1. Classical theory and the
properties of ghost states'', 
{\em Ann. Inst. Henri Poincar\'e}, {\bf 28} (1978)
225~; ``Nonsemisimple gauge models: 2. Renormalization'', {\em
Ann. Inst. Henri Poincar\'e}, {\bf 28} (1978) 255.

\bibitem{GW} J. Gomis and S. Weinberg, ``Are nonrenormalizable gauge
theories renormalizable?'', {\em Nucl. Phys.} {\bf
B469} (1996) 473, hep-th/951
0087.

\bibitem{HS1} T. Hurth and K. Skenderis, ``Quantum Noether
method'', {\em Nucl. Phys.} {\bf B541} (1999) 566,  hep-th/9803030.

\bibitem{HS2} T. Hurth and K. Skenderis, ``The Quantum Noether
condition in terms of interacting fields'', in {\it 
New Developments in Quantum Field Theory}, eds. P. Breitenlohner,
D. Maison and J. Wess (Springer, Berlin, to appear), hep-th/9811231.

\bibitem{Henn1} M. Henneaux, 
``On the algebraic structure of the BRST symmetry'', in 
{\em NATO Advanced Summer Institute and Banff Summer School in
Theoretical Physics on Physics, Geometry and Topology, Banff,
Canada, Aug 14-25, 1989},  H.C. Lee ed. (Plenum Press, New York, 1990)~; 
M. Henneaux and C. Teitelboim,
{\em Quantization of Gauge Systems} (Princeton University Press,
Princeton, 1992).

\bibitem{Henn2} M. Henneaux, 
``On the gauge-fixed BRST cohomology'',
{\em Phys. Lett.} {\bf B 367} (1996) 163,
hep-th/9510116. 

\bibitem{CurciF} G. Curci and R. Ferrari, ``On a class of Lagrangian
models for massive and massless Yang-Mills fields'', 
{\em Nuovo Cim.} {\bf 32 A}
(1976) 151~; 
``The unitarity problem and the zero - mass limit for a model of
massive Yang-Mills theory'', 
{\em Nuovo Cim.}
{\bf 35 A} (1976) 1.

\bibitem{Ojima} I. Ojima, ``Comments on massive and massless
Yang-Mills Lagrangians with a quartic coupling of Faddeev--Popov
ghosts'', {\em Z. Phys.} {\bf C 13} (1982) 173.

\bibitem{Kostasetal} J. de Boer, K. Skenderis, P. van Nieuwenhuizen
and A. Waldron, ``On the renormalizability and unitarity of the 
Curci-Ferrari model for massive vector bosons'',
{\em Phys. Lett.} {\bf B367} (1996) 175, hep-th/9510167.

\bibitem{Hurth1} T. Hurth, ``Nonabelian gauge symmetry in the causal
Epstein--Glaser approach'', {\em Int. J. Mod. Phys.} {\bf A12} (1997)
4461, hep-th/9511139.

\bibitem{Dragon} N. Dragon, T. Hurth and P. van Nieuwenhuizen,
``Polynomial form of the Stueckelberg model'', {\em
Nucl. Phys. Proc. Suppl.} {\bf B56} (1997) 318, 
hep-th/9703017. 

\bibitem{Hurth2} T. Hurth, ``Higgs-free massive nonabelian gauge
theories'', {\em Helv. Phys. Act.} {\bf 70}
(1997) 406, hep-th/9511176.

\bibitem{Brandt} F. Brandt, ``Deformations of global symmetries in the
extended antifield formalism'', 
{\em J. Math. Phys.} {\bf 40} (1999) 1023,
hep-th/9804153.

\bibitem{BBH1} G. Barnich, F. Brandt and M. Henneaux, ``Local
BRST cohomology in the antifield formalism. I. General theorems'', {\em
Commun. Math. Phys.} {\bf 174} (1995) 57, hep-th/9405109.

\bibitem{Vino} A.M. Vinogradov, ``On the algebra-geometric foundations
of Lagrangian field theory", {\em  Sov. Math. Dokl.} 
{\bf 18} (1977) 1200, ``A spectral sequence associated with a
nonlinear differential equation and algebra-geometric foundations
of Lagrangian field theory with constraints", {\em  Sov. Math. Dokl.}
{\bf 19} (1978) 1028; ``The theory of higher infinitesimal
symmetries of non linear partial differential equations",
{\em  Sov. Math. Dokl.} {\bf 20} (1979) 985.

\bibitem{Bryant} R.L. Bryant and P.A. Griffiths, {\em 
Characteristic Cohomology of Differential
Systems (I): General Theory} (Duke University Mathematics 
Preprint Series, volume 1993 n$^0$1, January 1993).

\bibitem{FischH} J.M.L. Fisch and M. Henneaux, ``Homological
perturbation theory and the algebraic structure of the antifield -
antibracket formalism for gauge theories'', {\em Commun. Math.
Phys.} {\bf 128} (1990) 627.

\bibitem{Henn0} M. Henneaux, ``Lectures on the antifield-BRST
formalism  for gauge theories'', {\em Nucl. Phys. B (Proc. Suppl.)}
{\bf 18A}
(1990) 47.

\bibitem{BaHe} G. Barnich and M. Henneaux, ``Consistent couplings
between fields with a gauge freedom and deformations of the master
equation'', {\em Phys. Lett.}
{\bf B311} (1993) 123, hep-th/9304057.

\bibitem{BS} N.N. Bogoliubov and D.V. Shirkov, {\it Introduction
to the Theory of Quantized Fields} (Interscience, New York, 1959).

\bibitem{EG0} H. Epstein and  V. Glaser, 
``Le r\^ole de la localit\'e dans la renormalisation perturbative 
en th\'eorie quantique des champs'', in {\em Statistical Mechanics
and Quantum Field Theory}, Proceedings of the 1970 Summer School of 
Les Houches, eds. C. DeWitt and R. Stora (Gordon and Breach, New York,
1971);  ``The r\^ole of locality
in perturbation theory'', {\em Ann. Inst. Poincar\'{e}} {\bf 29} 
(1973) 211. 

\bibitem{EG2} H. Epstein and V. Glaser, ``Adiabatic limit in perturbation 
theory'', in G. Velo, A.S. Wightman (eds.),  {\it
Renormalization Theory} (D. Reidel Publishing Company, Dordrecht 1976), p.
193;
O. Piguet and  A. Rouet, ``Symmetries in Perturbative  Quantum Field
Theory'', Phys. Rep. {\bf 76} (1981) 1;
H. Epstein, V. Glaser and R. Stora, ``General properties
of the $n$-point functions in local quantum field theory'', in J. Bros, 
D. Jagolnitzer (eds.), Les Houches Proceedings 1975; 
G. Popineau and R. Stora, ``A pedagogical remark on 
the main theorem of perturbative renormalization theory'', unpublished;
R. Stora, ``Differential Algebras'', ETH-lectures (1993) unpublished;
G. Scharf, ``Finite Quantum Electrodynamics'', 
Text and Monographs in Physics (Springer, Berlin, 1995);
T.~Hurth, ``NonAbelian gauge theories: the causal approach,''
{\em Ann. Phys.} {\bf 244} (1995) 340,
hep-th/9411080.

\bibitem{fred}
M.~D\"utsch and K.~Fredenhagen,
``A local (perturbative) construction of observables in gauge theories:  The example of QED,''
Commun.\ Math.\ Phys.\ {\bf 203} (1999) 71,
hep-th/9807078.

\bibitem{HS3} T. Hurth and K. Skenderis, ``Analysis of anomalies in
the Quantum Noether method'', in preparation.

\bibitem{Henneaux:1991rx}
M.~Henneaux, ``Space-time locality of the BRST formalism'', 
{\em Commun. Math.
 Phys.} {\bf 140} (1991) 1.
\end{thebibliography}
\end{document}